\documentclass[10pt,conference]{IEEEtran}
\usepackage[top=0.77in, bottom=1.03in, left=1in, right=1in]{geometry}

\IEEEoverridecommandlockouts

\IEEEoverridecommandlockouts

\usepackage{cite}
\usepackage{amsmath,amssymb,amsfonts}
\usepackage{algorithmic}
\usepackage{graphicx}
\usepackage{textcomp}
\usepackage{xcolor}
\usepackage{breqn}
\usepackage{algorithm,algorithmic}
\usepackage{algorithm,algorithmic,amsbsy,amsm     
 ath,amssymb,epsfig,bbm,mathrsfs,fancyhdr,fancyvrb,url}
\usepackage{amssymb}
 
\usepackage{amsthm}
\usepackage{adjustbox,lipsum}
\usepackage{amsfonts}
\usepackage{epsfig}
\usepackage{amssymb}
\usepackage{amsmath}
\usepackage{cite}
\usepackage{subfigure}
\usepackage{multirow}
\usepackage{rotating}
\usepackage{graphicx}
\usepackage{tabularx}
\usepackage{array}
\usepackage{anyfontsize}
\usepackage{color,soul}
\usepackage{graphicx}
\usepackage{blindtext}
\usepackage{authblk}

\usepackage{amsmath}
\usepackage{mathrsfs}

\usepackage{stfloats}
\usepackage{verbatim}
\usepackage{balance}

\usepackage{algorithm}
\floatname{algorithm}{Technique}       

\def\BibTeX{{\rm B\kern-.05em{\sc i\kern-.025em b}\kern-.08em
    T\kern-.1667em\lower.7ex\hbox{E}\kern-.125emX}}

\title{Exploring the Impact of HAPS-RIS on UAV-Based Networks: \\a Novel Network Architecture}

\author{}
\author{Arman Azizi*, Mustafa A. Kishk†, Arman Farhang* \\ *Department of Electronic and Electrical Engineering, Trinity College Dublin, Dublin,   
Ireland \\ † Department of Electrical and Electronic Engineering, Maynooth University, Maynooth, Ireland\\Emails: azizia@tcd.ie, mustafa.kishk@mu.ie, arman.farhang@tcd.ie
\thanks{\textcolor{black}{This publication has emanated from research conducted with the financial support of Taighde Éireann – Research
Ireland under Grant number 18/CRT/6222 and Grant number 13/RC/2077}\textunderscore\textcolor{black}{P2. For the purpose of Open Access, the author has applied
a CC BY public copyright licence to any Author Accepted Manuscript version arising from this
submission.} }
}
\begin{document}
\maketitle
\begin{abstract} 
In this paper, we propose a novel network architecture where two types of aerial infrastructures together with a ground station provide connectivity to a remote area. A high altitude platform station (HAPS) is equipped with reconfigurable intelligent surface (RIS), called HAPS-RIS, to be exploited to assist the unmanned aerial vehicle (UAV)-based wireless networks. A key challenge in such networks is the restricted number of UAVs, which limits full coverage and leaves some users unsupported. To tackle this issue, we propose a hierarchical bilevel optimization framework including a leader and a follower problem. The users served by HAPS-RIS are in a zone called the HAPS-RIS zone and the users served by the UAVs are in another zone called the UAV zone. In the leader problem, the goal is to establish the zone boundary and practical RIS phase shift design that maximizes the number of users covered by HAPS-RIS while ensuring that users in this zone meet their rate requirements. \textcolor{black}{This is achieved through our proposed practical relaxation method and the proposed dynamic radius-based zone association with RIS clustering (DyRaZARC) technique.} The follower problem focuses on minimizing the number of UAVs required, ensuring that the rate requirements of the users in the UAV zone are met. \textcolor{black}{This is addressed through our proposed novel geometry-informed machine learning method, called k-means adaptive dynamic UAV selection (KADUS) technique}. Our study reveals that increasing the number of RIS elements significantly decreases the number of required UAVs.

\end{abstract}
\begin{IEEEkeywords}
RIS, NTNs, HAPS, Network Architecture, Geometry-informed Machine Learning
\end{IEEEkeywords}
\section{Introduction}
\IEEEPARstart{A} primary aim of future wireless networks is to ensure universal connectivity, offering reliable access to everyone and everything, anywhere and anytime, at an affordable price. This includes extending coverage to remote areas and the areas where establishing terrestrial infrastructure is not feasible or affordable \cite{3GPP release}. To achieve this aim, unmanned aerial vehicles (UAVs), can play a vital role, operating as aerial base stations in regions where there is no cellular infrastructure or the areas where building a cellular infrastructure is very expensive. 
This includes scenarios like remote regions and disaster-stricken areas where ground communications may be destroyed, or any area where it is not affordable for network operators to establish terrestrial infrastructure \cite{azizi2019joint, azizi2019profit, saeedi2018throughput}. However, one of the main challenges in the UAV-assisted networks is the limitation on the number of UAVs that can be deployed, which leads to limitation on the coverage.
Deploying a network of multiple UAVs to provide full coverage over a specific area requires significant investment by network providers. Therefore, it is essential to minimize the number of UAVs to improve the practical deployment of multi-UAV networks \cite{sabzehali2022optimizing}.
With the limitation on the number of UAVs, full coverage cannot be achieved and a portion of users are left unsupported. One solution to tackle this issue is to use other types of non-terrestrial infrastructures as transmitters or relays. However, this approach is neither low-cost nor low-complexity, as it requires the implementation of multiple antennas on the non-terrestrial networks (NTN) \cite{NTN-RIS-alouini}. Another option is to utilize available terrestrial infrastructures despite their far distance to support the UAV-based network. This can be done by connecting them to NTN equipped with a reconfigurable intelligent surface (RIS).
This system is called non-terrestrial (NT)-RIS, a smart reflective layer acting as an intelligent intermediary for signal reflection. Exploiting NT-RIS has lower cost and less complexity than the aforementioned approach, using other types of non-terrestrial infrastructures \cite{NTN-RIS-alouini}. A considerable amount of research has been dedicated to exploring the advantages of implementing NT-RIS in wireless networks, in recent years, see \cite{NTN-RIS-alouini, NTN-RIS-jamalipour, Halim-HAPS-UAV-SAT, Halim-link-budget} and the related references therein. High altitude platform station (HAPS)-RIS stands out as a promising option for NT-RIS deployment, offering advantages over other NT-RIS like satellite-RIS and UAV-RIS \cite{Halim-HAPS-UAV-SAT, Halim-link-budget}.
\textcolor{black}{Accordingly, the main research question that arises is ``\textit{Can HAPS-RIS effectively address the mentioned challenge which is the limitation on the number of UAVs?}"}

In \cite{Halim-link-budget}, the link budget of the RIS-enabled NTN are studied and location of the HAPS-RIS is optimized. This study finds that HAPS, with its large area for RIS deployment, outperforms other aerial platforms like UAVs and low earth orbit (LEO) satellites in terms of signal strength. 
In \cite{alfattani2022beyond}, the authors present a scenario where HAPS-RIS is exploited to assist the network supported by terrestrial base stations and improve its service in urban areas. The users which are not covered with the terrestrial base stations, can be connected to the dedicated control station (CS) through the HAPS-RIS system. 
The paper introduces optimization problems including throughput maximization and minimizing the number of RIS elements by designing the RIS phase shifts and the power allocation of the CS. In \cite{alfattani2023resource}, the authors propose resource efficient approach, for the same scenario as \cite{alfattani2022beyond}, that aims to connect the maximum number of users while also minimizing the overall power usage of the CS and the RIS.
In \cite{azizi2023ris}, a multi-objective optimization problem is proposed where two unconnected infrastructures are connected though an aerodynamic HAPS-RIS. A closed-form solution for the RIS phase shifts are obtained to maximize the cascade channel gain, eliminate Doppler spread, and minimize the upper limit of delay spread. 

While there has been considerable progress in exploiting HAPS-RIS to improve wireless networks, a critical gap remains in applying this technology to UAV-assisted networks. To the best of our knowledge there is no work which studies the integration of HAPS-RIS technology and UAV-assisted networks where they serve as aerial base stations. In summary, this paper addresses the aforementioned gap in the literature by highlighting the following contributions.
\begin{itemize}
\item  We introduce a \textit{novel network architecture} where a supportive network exploiting HAPS-RIS is established to address a key challenge in UAV-based networks, reducing the number of required UAVs.
\item We mathematically model the heterogeneous zones, i.e., UAV zone and HAPS-RIS zone for user clustering and associate zone users with their allocated infrastructure, i.e., UAV or HAPS-RIS.
\item We propose a novel hierarchical bilevel optimization problem, including a leader and a follower problem. In the leader one, we obtain the zone boundary and \textcolor{black}{a practical design for the RIS phase shifts} to maximize the number of users supported by HAPS-RIS while satisfying the rate requirements for each user in the HAPS-RIS zone. \textcolor{black}{We practically relax and redesign the leader problem into an optimization problem with the lower computational complexity, and solve it with the proposed dynamic radius-based zone association with RIS clustering (DyRaZARC) technique.} In the follower optimization problem, we aim to minimize the number of UAVs while meeting the rate requirement for each user in the UAV zone. \textcolor{black}{To solve the follower problem, we propose a novel geometry-informed machine learning method, called $k$-means adaptive dynamic UAV selection (KADUS) technique}. Our findings demonstrate that enhancing the number of RIS elements can significantly reduce the required number of UAVs, revealing a key strategy for optimizing aerial operations.
\end{itemize}



   
 \section{Proposed Network Architecture}
\begin{figure}
\centering
\includegraphics[scale=0.43]{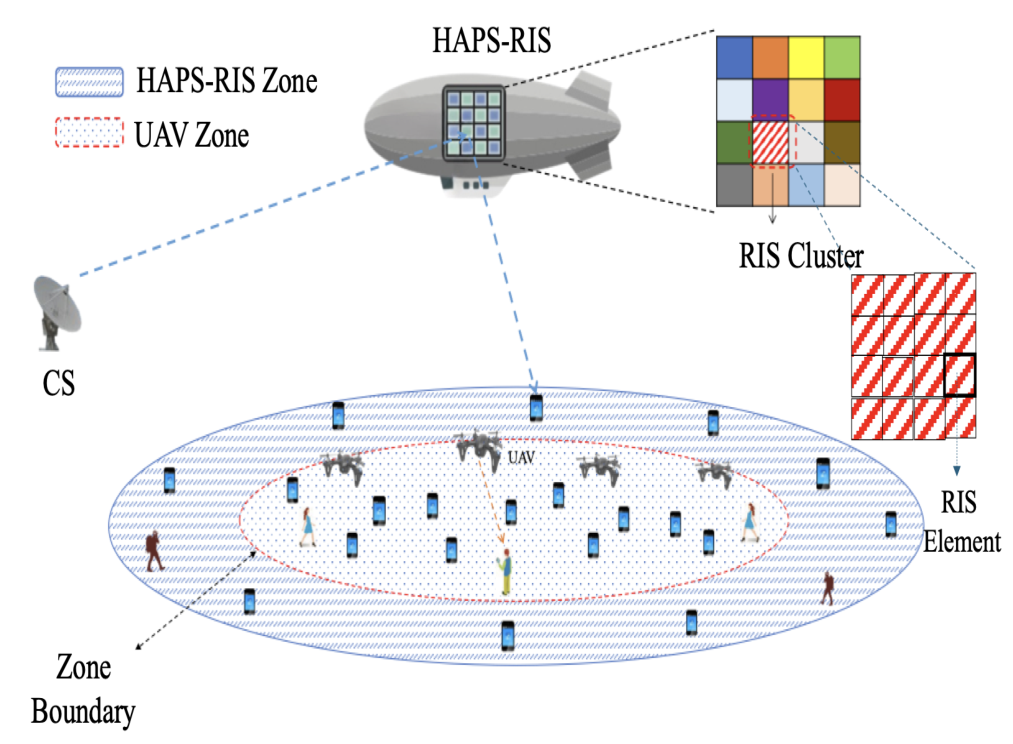}
    \caption{The Proposed Network Architecture}
    \label{sys2}
\end{figure}

We consider a downlink transmission scenario, shown in Fig. \ref{sys2}, where heterogeneous non-terrestrial infrastructures are exploited to bring connectivity to a remote area. The set of \textcolor{black}{stationary} users $\mathcal{I}$ is defined as $\mathcal{I}$=\{1, \dots, $I$\}, where $i \in \mathcal{I}$ represents an individual user indexed by $i$. The users are uniformly distributed in a circular area where there is no nearby terrestrial infrastructure. 
The set of UAVs $\mathcal{J}$ is defined as $\mathcal{J}$ = \{1, \dots, $J$\}, where $j \in \mathcal{J}$ represents an individual UAV indexed by $j$. \textcolor{black}{In this scenario, the UAVs remain stationary, hovering in place to provide service to the fixed users.} The aim is minimizing the number of required UAVs with exploiting terrestrial CS and HAPS-RIS while providing full coverage with meeting the minimum rate requirements. In HAPS-RIS scenarios, CS uses high gain directional antenna, creating a dominated line of sight (LoS) connection with HAPS \cite{Halim-HAPS-UAV-SAT,Halim-link-budget, alfattani2022beyond, alfattani2023resource,azizi2023ris}.
We consider UAVs and CS operate independently to provide connectivity and they are not connected to each other. All UAVs can be connected to the core network via a backhaul link. Due to heavy blockage, there is no direct link between the users and the CS; communication instead occurs through a cascaded channel through HAPS-RIS. 
 The set of RIS elements $\mathcal{M}$ is defined as $\mathcal{M}$ = \{1, \dots, $M$\}, where $m \in \mathcal{M}$ represents the RIS element indexed by $m$.
 Users can be connected to the network either through the UAV system or the HAPS-RIS setup, but not through both at the same time. The coverage area is a circle with the radius $R_0$ with the centre of $(x_0,y_0, z_0)$, the CS location is $(x^{\rm{CS}},y^{\rm{CS}},z^{\rm{CS}})$, the location of the HAPS is $(x^{\rm{HAPS} },y^{\rm{HAPS} },z^{\rm{HAPS}})$, and the location of the UAV $j$ is $(x_{j}^{\rm{UAV} },y_{j}^{\rm{UAV} },z_{j}^{\rm{UAV}})$ in the Cartesian coordinate system. 
As can be seen in Fig. \ref{sys2}, there are two zones in the coverage area, i.e., the inner circle with the radius $R$ and the ring around it. The \textit{UAV zone} is the area where the users are covered by UAVs and the \textit{HAPS-RIS zone} is the ring area between the inner and outer circles, in which the users are covered by HAPS-RIS. The border which separates the UAV zone and the HAPS-RIS zone is called \textit{zone boundary}.
$\mathcal{B}$ is the set which includes the users in the UAV zone. Thus, $\mathcal{C}$=$\mathcal{I-B}$ is the set which includes users in the HAPS-RIS zone.
The set of equally spaced orthogonal subcarriers $\mathscr{L}$ is defined as $\mathscr{L}$=\{1, \dots, $L^{\rm{tot}}$\}, where $l \in \mathscr{L}$ represents the subcarrier indexed by $l$. 
Let us consider the available bandwidth of $BW$ which is divided into $L^{\rm tot}$. Therefore, the bandwidth allocated to each orthogonal subcarrier is $B_{l}=\frac{BW}{L^{\rm{tot}}}$.
\section{Problem Formulation and Proposed Solution Techniques}
We propose a hierarchical bilevel optimization problem, including the leader problem \textbf{OP 1} and the follower problem \textbf{OP 2}. In \textbf{OP 1}, we determine the zone boundary in order to maximize the number of users in the HAPS-RIS zone while satisfying the rate requirement for each user. We consider \textbf{OP 1} as the leader problem, because the aim is to fully utilize the resources of HAPS-RIS system to assist the UAV-based network by maximizing the number of users in the HAPS-RIS zone. Accordingly, the number of users in the UAV zone is minimized, and hence, less number of UAVs is required.
After solving \textbf{OP 1}, in the follower problem \textbf{OP 2}, we obtain the minimum number of UAVs subject to the user rate requirements for the users in the UAV zone.
\subsection{Proposed HAPS-RIS Zone Design}
In the HAPS-RIS zone,
$h^{\rm{CS} }_{im}$ is the cascade channel gain between user $i$-RIS and RIS-CS through RIS element $m$.
\begin{equation}
 \begin{array}{l}
          h^{\rm{CS} }_{im}=\\
    \sqrt{G^{\rm{CS} }G_{i}\left( \mathcal{L}^{\rm{CS-RIS} }_{m}\right)^{-1}  \left( \mathcal{L}^{\rm{RIS-user} }_{im}\right)^{-1}},  \forall m, \forall i\in\mathcal{C},
    \end{array}
\end{equation}

where $G^{\rm{CS}}$ is the antenna gain of the CS and $G_{i}$ is the antenna gain of the user $i$. 
 $(x^{\rm{CS}},y^{\rm{CS} },z^{\rm{CS} })$, 
 $\left( x_{i},y_{i},z_{i}\right)$,
 $(x^{\rm{HAPS} },y^{\rm{HAPS} },z^{\rm{HAPS}})$, $\left( x^{\rm{RIS} }_{m},y^{\rm{RIS} }_{m},z^{\rm{RIS} }_{m}\right)$ are the CS, user $i$,  HAPS, and RIS element $m$, coordinates, respectively.
 Using the Friis model, \cite{Friis}, the path loss between CS and RIS element $m$, $\forall m$, can be formulated as

\begin{equation}
   \begin{array}{l}
 \mathcal{L}^{\rm{CS} ,\rm{RIS} }_{m}=(\frac{4\pi f_{\rm c}}{c} )^{2}\times\\
       ( ( x^{\rm{RIS} }_{m}-x^{\rm{CS} })^{2}+(y^{\rm{RIS} }_{m}-y^{\rm{CS} })^{2}+(z^{\rm{RIS} }_{m}-z^{\rm{CS}})^{2} ),  
    \end{array}
\end{equation}
and the path loss between RIS element $m$ and the user $i$, $\forall m, \forall i \in \mathcal{C}$, can be formulated as 
\begin{equation}
   \begin{array}{l}
\mathcal{L}^{\rm{RIS}, \rm{user}}_{im} = \left( \frac{4\pi f_{\rm{c}}}{c} \right)^{2} \times \\
 ( (x^{\rm{RIS}}_{m} - x_{i})^{2} + (y^{\rm{RIS}}_{m} - y_{i})^{2} + (z^{\rm{RIS}}_{m} - z_{i})^{2} ), 
    \end{array}
\end{equation}
where $f_{\rm c}$ is the carrier frequency and $c$ is the speed of wave propagation which is considered to be equal to the speed of light. 
The communication rate of the user $i$ received from the CS through the HAPS-RIS can be obtained as
\begin{equation}\label{rateHAPS}
    r^{\rm{CS}}_{i}= \sum \limits ^{L^{\rm{CS}}}_{l=1} B_{l}\log_{2}\left( 1+\gamma^{l,\rm{CS} }_{i} \right) , \forall i\in\mathcal{C},
\end{equation}
\textcolor{black}{where $L^{\rm{CS}}$=$\frac{L^{\rm{tot}}}{2}$ is available subcarriers}, $\gamma^{l,\rm{CS} }_{i}$, the signal to noise ratio (SNR) of the received signal at user $i$ and subcarrier $l$, can be obtained as 
\begin{equation}\label{SNRHAPS}
   \gamma^{l,\rm{CS} }_{i}=\frac{{P^{l,\rm{CS} }}\left|\sum \limits_{m=1}^{M}\rho^{l,\rm{CS} }_{im} h^{\rm{CS} }_{im}\vartheta_{im} \right|^{2}  }{N_{0}B_{l}}, \forall i\in\mathcal{C}, \forall l,
 \end{equation}
 where $M$ is the total number of RIS elements, $N_{0}$ is the noise power spectral density, and $P^{l,\rm{CS} }$ is the CS transmit power allocated uniformly to the subcarrier $l$. $\rho^{l,\rm{CS} }_{im}$ is the binary variable which demonstrates the user $i$ is assigned to RIS element $m$ over subcarrier $l$.

 Considering $\vartheta_{im}$ the reflection coefficient of the RIS element $m$ corresponding to the user $i$ as 
\begin{equation}
\vartheta_{im} =\mu_{m} e^{-j(\phi_{m} -\xi^{\rm{CS} }_{m} -\omega_{im} )}, \forall i\in\mathcal{C},m,
\end{equation}
where $\phi_{m}$ is the RIS phase shift for the element $m$. $\xi^{\rm{CS} }_{m}$ is the corresponding phase between RIS element $m$ and the CS. $\omega_{im}$ is the corresponding phase between RIS element $m$ and the user $i$. $\mu_{m}$ is the reflection loss corresponding to the RIS element $m$. \\

 \textcolor{black}{\textbf{\textit{Remark 1}}}:  Users in the HAPS-RIS zone are assumed to be orthogonal to each other and to the users of the UAV zone. Hence, there is no inter-user interference.\\
 
\textcolor{black}{\textbf{\textit{Definition 1}}: Similar to the approach in \cite{alfattani2023resource}, where the justification is elaborated in detail, we consider each user in the HAPS-RIS zone is allocated to its own group of subcarriers and RIS elements. The group of RIS elements which are allocated to a user is called \textit{RIS cluster}. Given this assumption, $\rho^{l,\rm{CS} }_{im}$ can be obtained based on the value of $R$, the radius of the inner circle, which determines the zone boundary, \textcolor{black}{elaborated in the proposed DyRaZARC Technique.}} \textcolor{black}{Taking all these factors into account, the leader optimization problem can be formulated as} \\
\begin{subequations}
    \begin{equation}
    \hspace{-4cm}
\textcolor{black}{\textbf{OP 1}:~~~ \max_{{R},\vartheta_{im}} \  U^{\rm{HAPS}}}
\end{equation}
\begin{equation}\label{rateconst}
    \hspace{-1.3cm}
 \textcolor{black}{{{\rm{s}}.{\rm{t}}.:} ~~ r^{\rm{CS} }_{i}(R,\vartheta_{im})  \geq r^{\rm{user} }_{0}, \forall i\in\mathcal{C},}
\end{equation}
\end{subequations}
\textcolor{black}{where $U^{\rm{HAPS}}$ is the number of users covered by HAPS-RIS in the HAPS-RIS zone and $r^{\rm{user} }_{0}$ is the minimum rate requirement for each user.}\\

 \textcolor{black}{\textbf{\textit{Proposition 1}}: \textbf{OP 1} can be relaxed into }
\begin{subequations}
    \begin{equation}
    \hspace{-4cm}
\textcolor{black}{\textbf{OP 1-a}:~~~ \max_{{R}} \  U^{\rm{HAPS}}}
\end{equation}
\begin{equation}\label{ratecons2}
    \hspace{-1.3cm}
\textcolor{black}{ {{\rm{s}}.{\rm{t}}.:} ~~ r^{\rm{CS} }_{i}(R)  \geq r^{\rm{user} }_{0}, \forall i\in\mathcal{C},}
\end{equation}
\end{subequations}
\textcolor{black}{where $\phi^{\ast }_{m}= \xi_{0} +\omega_{0} +2k\pi, ~\forall m$ and $k$ can be an integer value.}
\begin{proof} 
In the proposed geometrical case, the dimensions of RIS can be negligible compared to the altitude of HAPS, as it does not have much effect on the performance. Hence, without loss of generality, we relax the location of the RIS elements as 
\begin{equation}
    \begin{matrix}\left( x^{\rm{RIS} }_{m},y^{\rm{RIS} }_{m},z^{\rm{RIS} }_{m}\right)  \approx (x^{\rm{HAPS} },y^{\rm{HAPS} },z^{\rm{HAPS} }),&\forall m\end{matrix},
\end{equation}
the dimensions of the RIS and the coverage area is much less than the range of the distance between HAPS-RIS and the terrestrial components.\footnote{\textcolor{black}{The dimensions of the RIS are in the range of 100 meters while the distance between the HAPS-RIS and the terrestrial network is in the range of more than 20 kilometers.}} Accordingly, without loss of generality, we can relax the corresponding phases to fixed values as $\xi^{\rm{CS} }_{m} \approx \xi_{0}, \forall m $ and $\omega_{im} \approx \omega_{0}, \forall i\in\mathcal{C}, \forall m,$ so the reflection gain of RIS element $m$ can be relaxed as $\vartheta_{im} \approx {\tilde\vartheta_{m}}, \forall i\in\mathcal{C}, \forall m$ where
\begin{equation}\label{relaxedreflection}
{\tilde\vartheta_{m}}= \mu_{m} e^{-j(\phi_{m} -\xi_{0} -\omega_{0} )}\  \  \  \forall m.
\end{equation}
\textcolor{black}{As \eqref{relaxedreflection} is independent from user index, $r^{\rm{CS} }_{i}(R,\tilde\vartheta_{m})$ can be relaxed to its upper bound by considering $\phi^{\ast }_{m} =$$\arg \max_{\phi_{m}}$$ {\tilde{\vartheta }_{m}}=\xi_{0} +\omega_{0} +2k\pi, \forall m$. Accordingly, \eqref{rateconst} is relaxed into \eqref{ratecons2} and hence \textbf{OP 1} is relaxed into \textbf{OP 1-a}.  
\textcolor{black}{The proposed technique significantly reduces the computational complexity by eliminating the need to individually optimize the phase shifts for each RIS element. Given a massive number of RIS elements in  HAPS-RIS scenarios, ranging from $10^5$ to $10^8$, see Section. \ref{simul}, the proposed design can pave the way for scalability and enables practical implementation.}}
\end{proof}

\textbf{\textit{Proposition 2}}: \textbf{OP 1-a} can be transformed as
\begin{subequations}
    \begin{equation}
    \hspace{-4cm}
\textbf{OP 1-b}:~~~ \min_{R} \  R^2
\end{equation}
\begin{equation}
    \hspace{-1cm}
 {{\rm{s}}.{\rm{t}}.:} ~~ r^{\rm{CS} }_{i}(R)  \geq r^{\rm{user} }_{0}, \forall i\in\mathcal{C},
\end{equation}
\end{subequations}

\begin{proof} Maximizing $U^{\rm{HAPS}}$ is equivalent to maximizing the area of HAPS-RIS zone which is $\pi \left( R^{2}_{0}-R^{2}\right)$. $R_0$ is the outer circle radius and as it has a fixed value, maximizing $\pi \left( R^{2}_{0}-R^{2}\right)$ is equivalent to minimizing $\pi R^{2}$.
\end{proof}

\textcolor{black}{The solution of \textbf{OP 1-b}, i.e., \( R^* \), is obtained by the proposed DyRaZARC technique.} \textcolor{black}{In each iteration, the zone boundary and user sets are updated, and users in the HAPS-RIS zone are assigned to their respective RIS clusters using a random injective function, with an equal number of subcarriers allocated to each user. In the proposed geometric channel model for the HAPS-RIS zone, the channel gain is affected by the locations of the RIS clusters, the users, and the CS. Hence, the random allocation of RIS clusters is justified due to the geometric configuration of the system, where the RIS dimensions are negligible compared to the distances between the HAPS and the users. Furthermore, an injective function is adopted to ensure that each user in the HAPS-RIS zone is uniquely assigned to one of the available RIS clusters.} This process continues until the maximum number of users in the HAPS-RIS zone is achieved while satisfying the rate requirement. 
\begin{algorithm}[t]
\floatname{algorithm}{\textcolor{black}{The Proposed DyRaZARC Technique:}}
		\caption{Solving \textbf{OP 1-b} to Obtain the Zone Boundary}
		\label{alg:algorithm1}
		{\textbf{s1:}}~ Initialize $t=0$, $\delta=50$, $R^{*}=R_{0}$, and define the set $\mathcal{I}$=\{$i|\sqrt{x^{2}_{i}+y^{2}_{i}} \leqslant R_{0}$\} 
  
       {\textbf{s2:}}~ \textbf{for} $t=1$ to $T$ {\textbf{do}}
		\\{\textbf{s3:}}~ Define the set ${\mathcal{B}^{(t)}}$=\{$i'|\sqrt{x^{2}_{i'}+y^{2}_{i'}} \leqslant R^{(t)}$\} where $R^{(t)}=R_{0}-t\delta$. \textcolor{black}{$T$ is the upper limit of the iteration count}.
 \\ {\textbf{s4:}}~ Define the set $\mathcal{C}^{(t)}= \mathcal{I}-\mathcal{B}^{\left( t\right)}$ where the users in this area are supported by HAPS-RIS, and then calculate the number of the RIS elements per each cluster as $\left\lfloor\frac{M}{\left| \mathcal{C}^{(t)}\right|}\right\rfloor $.   
 
  {\textbf{s5:}}~ Map $\left| \mathcal{C}^{(t)}\right|$ clusters of RIS elements to $\left| \mathcal{C}^{(t)}\right|$  users based on a random injective function. Allocate $L^{(t)}_{\rm{Cluster}}=\lfloor\frac{L^{\rm{CS}}}{\left| \mathcal{C}^{(t)}\right|}\rfloor$ subcarriers to each RIS cluster. Then, uniformly allocate $L^{(t)}_{\rm{Cluster}}$ subcarriers to $\left\lfloor\frac{M}{\left| \mathcal{C}^{(t)}\right|}\right\rfloor$ RIS elements for each RIS cluster and calculate $\boldsymbol{\rho}(R^{(t)})$.
   
   {\textbf{s6:}}~\textbf{While} ~~$r^{\rm{CS} }_{i}\left( \boldsymbol{\rho} \left( R^{(t)}\right)  \right)  \geq r^{\rm{user} }_{0}$, $\forall i\in\mathcal{C}$ or $t < T$,  set $R^{*}=R^{(t)}$. 
 		\\ {\textbf{s7:}}~~~Set $t=t+1$ 
        \\~~~{\textbf{s8:}}~{\textbf{end for}}
		\\ {\textbf{s9:}}~ $R^{*}$ is the final solution of \textbf{OP 1-b}.
	\end{algorithm}

\subsection{Proposed UAV Zone Design} 
In the UAV zone, the communication rate of the user $i$ can be obtained as
\begin{equation}
    r^{\rm{UAV}}_{i}= \sum \limits^{J=N^{\rm{UAV}}}_{j=1} \sum \limits^{L^{\rm{UAV}}}_{l'=1} B_{l}\log_{2}\left( 1+\gamma^{l',\rm{UAV} }_{ij} \right), \forall i\in\mathcal{B},
\end{equation}
\textcolor{black}{where the available subcarriers are $L^{\rm{UAV}}$=$\frac{L^{\rm{tot}}}{2}$}. 
Hence, the signal to interference plus noise ratio (SINR) of the user $i$ served by UAV $j$ over subcarrier $l'$, $\forall i\in\mathcal{B}, \forall j, l'$, can be obtained as 
\begin{equation}
\begin{array}{l}
        \gamma^{l',\rm{UAV} }_{ij}=
        \frac{P^{l',\rm{UAV} }_{ij}\tilde{\rho}^{l',\rm{UAV} }_{ij} \left| h^{l',\rm{UAV} }_{ij} \right|^{2}  }{N_{0}B_{l}+\sum^{J}_{j^{\prime }\neq j} P^{l',\rm{UAV} }_{ij^{\prime }}\tilde{\rho}^{l',\rm{UAV} }_{ij^{\prime }} \left| h^{l',\rm{UAV} }_{ij^{\prime }} \right|^{2}  }, 
\end{array}
\end{equation} 
where $P^{l',\rm{UAV} }_{ij}$ is the transmit power of the UAV $j$ allocated to the subcarrier $l'$ and the user $i$. $h^{l',\rm{UAV} }_{ij}$ is the channel gain between UAV $j$ and user $i$ over subcarrier $l'$, which can be formulated as $    h^{\rm{UAV} }_{ij}=\sqrt{G^{\rm{UAV} }_{j}G_{i}\bar{\mathcal{L}}^{-1}_{ij}  }, \forall i\in\mathcal{B}, \forall j$. $G^{\rm{UAV} }_{j}$ is the antenna gain of the UAV $j$, $G_{i}$ is the antenna gain of the user $i$, and $\bar{\mathcal{L}}_{ij}$ is the average path loss, $\forall i\in\mathcal{B},  \forall j$, which can be obtained as \cite{al2014optimal}
\begin{equation}
 \begin{array}{l}
       \bar{\mathcal{L}}_{ij}=\\ P^{\rm{UAV} }_{ij,\rm{LoS} }\eta_{1} \left( \frac{4\pi f_{c}}{c} d_{ij}\right)^{\alpha }+ P^{\rm{UAV} }_{ij, \rm{NLoS}}\eta_{2} \left( \frac{4\pi f_{c}}{c} d_{ij}\right)^{\alpha},

    \end{array}
\end{equation}
where $\alpha$ is the path loss exponent, $\eta_{1}$ and $\eta_{2}$ are the excessive path losses 
corresponding to the LoS and non-LoS connections, respectively. The distance between the user $i$ and the UAV $j$, $\forall i\in\mathcal{B}, \forall j$, can be formulated as
\begin{equation}
    d_{ij}=\sqrt{\left( x_{i}-x^{\rm{} \rm{UAV} }_{j}\right)^{2}  +\left( y_{i}-y^{\rm{UAV} }_{j}\right)^{2}  +(z^{\rm{UAV} }_{j})^{2}}, 
\end{equation}
and $P^{\rm{UAV} }_{ij,\rm{} \rm{LoS} }$ is the LoS probability between the user $i$ and the UAV $j$ as
\begin{equation}
    P^{\rm{UAV} }_{ij,\rm{} \rm{LoS} }=\frac{1}{1+\psi e^{-\beta \left( \tilde{\theta}_{ij} -\psi \right)  }}, 
\end{equation}
where $\beta$ and $\psi$ are the constant values that depend on the environment type. $  P^{\rm{UAV} }_{ij,\rm{} \rm{LoS} }+P^{\rm{UAV} }_{ij,\rm{} \rm{NLoS} }=1, \forall i\in\mathcal{B}, \forall j,
$ and $ \tilde{\theta}_{ij} =\frac{180}{\pi } \arcsin ( \frac{z^{\rm{UAV} }_{j}}{d_{ij}}),  \forall i\in\mathcal{B}, \forall j$.\\

\textcolor{black}{\textbf{\textit{Remark 2}}}: In the UAV zone, we exploit orthogonal frequency division multiple access (OFDMA), and hence, there are multi UAV cells where there is inter-cell interference but no intra-cell interference.\\

\textcolor{black}{\textbf{\textit{Definition 2}}: $\boldsymbol{\tilde{\rho }}$ is obtained based on $N^{\rm{UAV}}$ via $k$-means clustering and satisfying subcarrier orthogonality in each UAV cluster. Considering all these factors, the follower optimization problem can be designed as}
\begin{subequations}
    \begin{equation}
    \hspace{-4cm}
\textbf{OP 2}:~~~ \min_{N^{\rm{UAV}}} \  N^{\rm{UAV}}
\end{equation}
\begin{equation}
    \begin{array}{l}
          {{\rm{s}}.{\rm{t}}.:} ~~ r^{\rm{UAV} }_{i}\left(N^{\text{UAV} } \right) \geq r^{\rm{user} }_{0},\  \forall i \in\mathcal{B}.
    \end{array}
\end{equation}
\end{subequations}
\textcolor{black}{The solution of \textbf{OP 2}, i.e., minimum number of required UAVs, is achieved by exploiting the proposed geometry-informed machine learning KADUS technique, for the given zone boundary which is obtained in \textbf{OP 1}. Unlike conventional $k$-means clustering, where the $k$-factor is a fixed input \cite{jain2010data}, the proposed KADUS technique considers it as a varying number of UAVs that adapts iteratively. This approach increases the network degrees of freedom.}
In each iteration, the users are associated to the UAVs using $k$-means clustering method while satisfying the subcarrier orthogonality and the rate requirement. The process continues until minimum number of UAVs is obtained.
\newline
\begin{algorithm}[t]
\floatname{algorithm}{\textcolor{black}{The Proposed Geometry-informed Machine Learning KADUS Technique:}}		\caption{Solving \textbf{OP 2}}
		\label{alg:algorithm2}
	 {\textbf{s1:}} Initialize  $\delta'$=1, $N^{\rm{UAV},(0)}$ and define the set ${\mathcal{B^{*}}}$=\{$i'|\sqrt{x^{2}_{i'}+y^{2}_{i'}} \leqslant R^{*}$\} where $R^{*}$ is the final solution of \textbf{OP 1-b}.
  
   {\textbf{s2:}}~ \textbf{for} $t=1$ to $T$ {\textbf{do}}
		
{\textbf{s3:}} Obtain $\boldsymbol{x}\left( N^{\text{UAV},(t)} \right)$, $\boldsymbol{y}\left( N^{\text{UAV},(t)} \right)$, and the user association part of $\boldsymbol{\tilde{\rho}}\left( N^{\text{UAV},(t)} \right)$ based on the $k$-means clustering method for the users as members of the set $\mathcal{B}^{*}$, where $N^{\rm{UAV},(t)} = N^{\rm{UAV},(0)} - (t-1)\delta'$.

  {\textbf{s4:}} Apply orthogonality check for subcarriers based on the constraint $\sum\limits^{I}_{i} \tilde\rho^{l'}_{ij} \leq 1\,\forall l',j$. Then calculate $\boldsymbol{\tilde\rho} \left( N^{\text{UAV},(t) }\right)$.
   
   {\textbf{s5:}}~\textbf{While} ~~ $t < T$ or
  $r^{\rm{UAV} }_{i}\left( \boldsymbol{\tilde\rho} \left( N^{\text{UAV},(t) }\right), \boldsymbol{x}\left( N^{\text{UAV},(t) }\right),\boldsymbol{y}\left( N^{\text{UAV},(t) }\right) \right) \geq r^{\rm{user} }_{0},\  \forall i \in\mathcal{B}$
  ,  set $N^{*}=N^{{\rm{UAV}},(t)}$.
		\\ {\textbf{s6:}}~~~Set $t=t+1$  
        \\{\textbf{s7:}}~{\textbf{end for}}
		\\ {\textbf{s8:}} $N^{*}$ is the final solution of \textbf{OP 2}.
\end{algorithm}

\textcolor{black}{\textbf{\textit{Corollary 1:}}
The solution of the proposed hierarchical bilevel optimization framework including the leader problem, \textbf{OP 1} which is defined in \textit{Definition 1}, and the follower one, \textbf{OP 2} which is defined in \textit{Definition 2},
is $(R, \boldsymbol{\vartheta}, N^{\text{UAV}})=(R^{*}, \boldsymbol{\vartheta^{*}}, N^{*})$. $\boldsymbol{\vartheta^{*}}$ can be obtained via \textit{Proposition 1} where the RIS phase shifts are practically designed in \textbf{OP 1-a}. With the given $\boldsymbol{\vartheta^{*}}$, $R^{*}$ can be obtained  via \textit{Proposition 2} and the proposed DyRaZARC technique. Then, with the given $R^{*}$, $N^{*}$ can be obtained by the proposed geometry-informed machine learning KADUS technique. }
\section{Numerical Evaluations}\label{simul}

\textcolor{black}{In the simulation setup, the selected values for the HAPS-RIS system are based on established models and findings in the relevant literature \cite{Halim-HAPS-UAV-SAT,Halim-link-budget}, while the values chosen for the UAVs align with the assumptions in \cite{al2014optimal,azizi2019profit}.}
We consider $I=20$ users which are spread in random locations following the uniform distribution inside a circle area with the center $(x_0,y_0, z_0)=(0,0,0)$ and the radius of $R_0=500$ meters. The CS is located at $(x^{\rm{CS}},y^{\rm{CS} },z^{\rm{CS} })=(-10000, 0, 1000)$ and the HAPS is located at $(x^{\rm{HAPS} },y^{\rm{HAPS} },z^{\rm{HAPS}})=(-5000,100,20000)$.
The CS antenna gain and each user antenna gain are considered $G^{\rm{CS}}=43.2$ dB, $G_{j}=0$ dB $\forall j$, $G_{i}=0$ dB $\forall i$, respectively. The system operates over a total bandwidth of $BW=100$ MHz, equally divided between a UAV-based network and a HAPS-RIS-assisted network. It uses a carrier frequency of $f_c=2$ GHz and $L=64$ subcarriers for communication. 
The reflection loss corresponding to each RIS element is considered $\mu_{m}=1, \forall m$. The noise power spectral density is $N_0=-174$ dBm/Hz, the path loss exponent is $\alpha=2$, and the excessive path loss coefficients for LoS and NLOS are $\eta_1=1$ and $\eta_2=31$, respectively. The constant values which depend on the environment type are considered $\psi=5$ and $\beta=0.5$. The CS transmit power is $40$ dBm except the results where its value is mentioned and the transmit power of each UAV is $20$ dBm. The results are averaged over a large number of simulation runs.

\begin{figure}[t]
\centering
\includegraphics[scale=0.35]{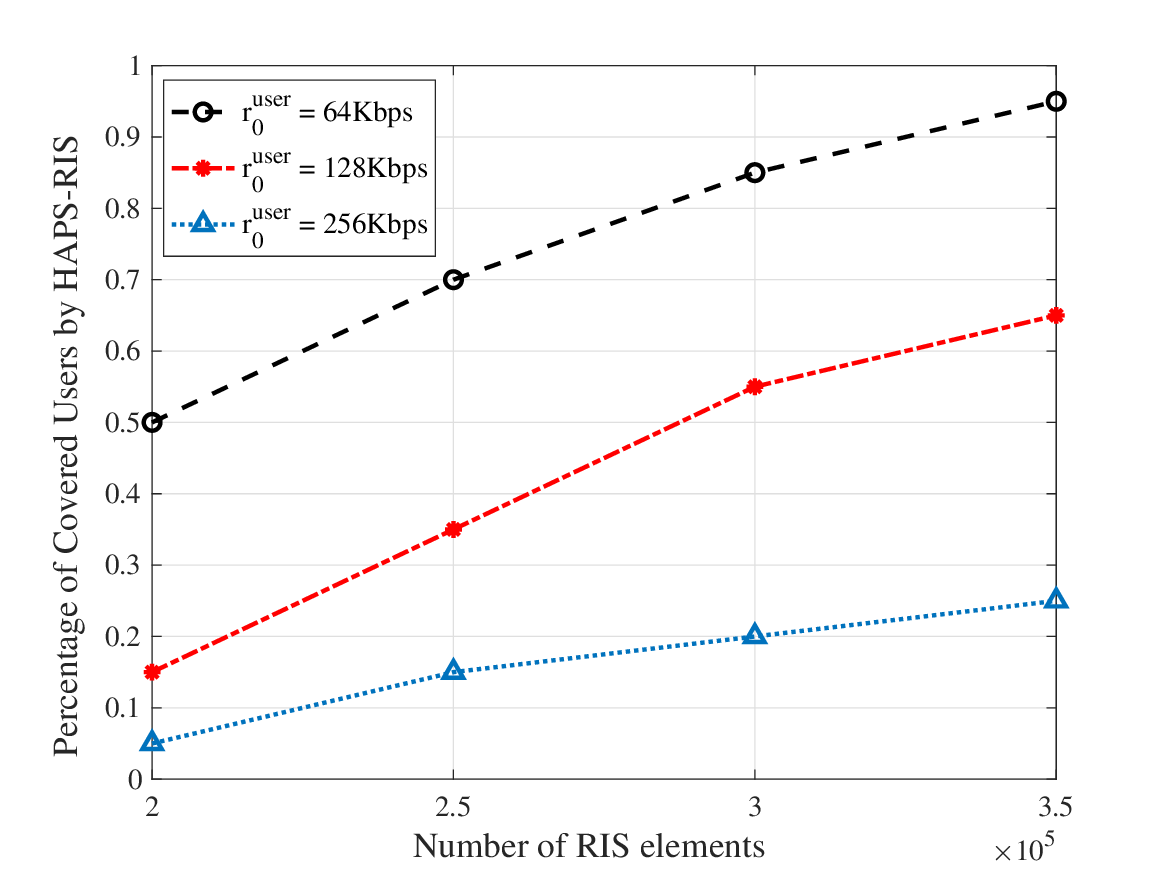}
    \caption{Percentage of covered users by HAPS-RIS versus number of RIS element for the rate requirements 64, 128, and 256 kbps when the CS transmit power is 40 dBm and only HAPS-RIS setup is in use.}
    \label{Me5CoveredUsers}
\end{figure}

\begin{figure}[t]
\centering
\includegraphics[scale=0.36]{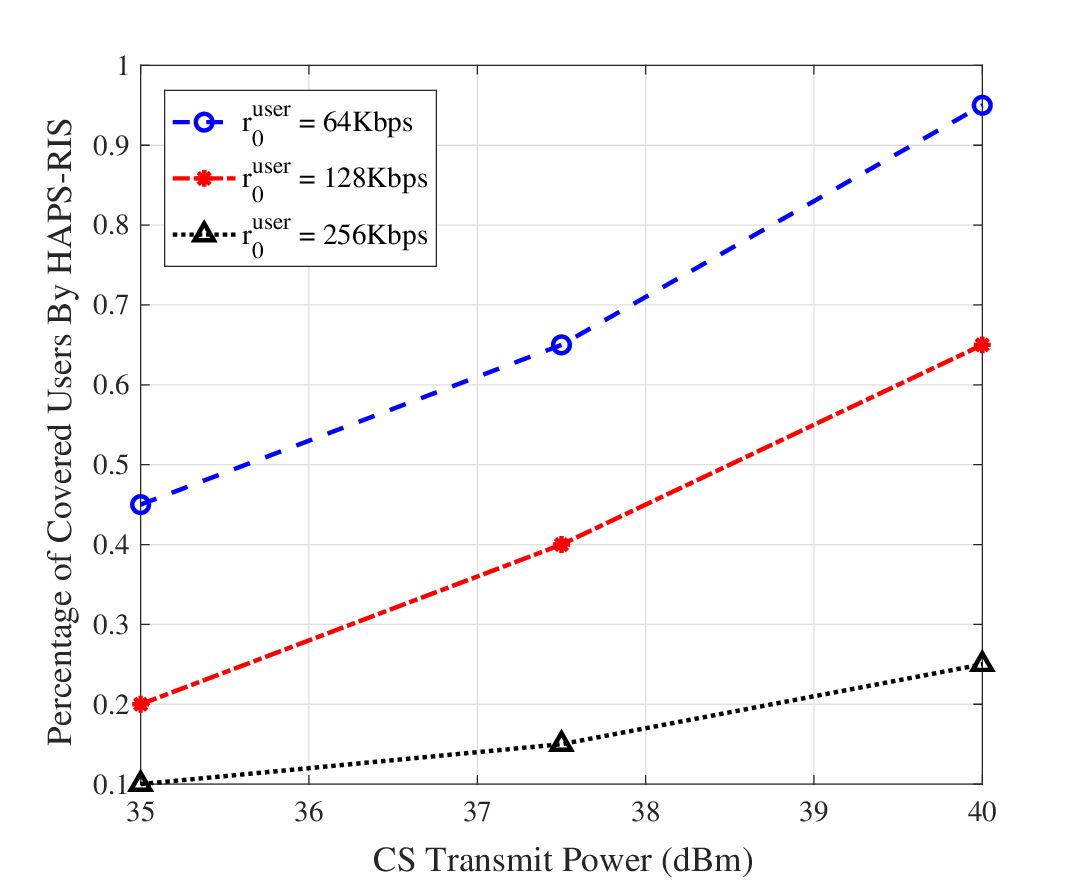}
    \caption{Percentage of covered users by HAPS-RIS versus CS Transmit power for the rate requirements 64, 128, and 256 kbps when the number of RIS elements are $3.5\times$$10^5$ and only HAPS-RIS setup is in use.}
    \label{CoveredUsersVSPowerMe5}
\end{figure}

\begin{figure}
\centering
  \includegraphics[scale=0.4]{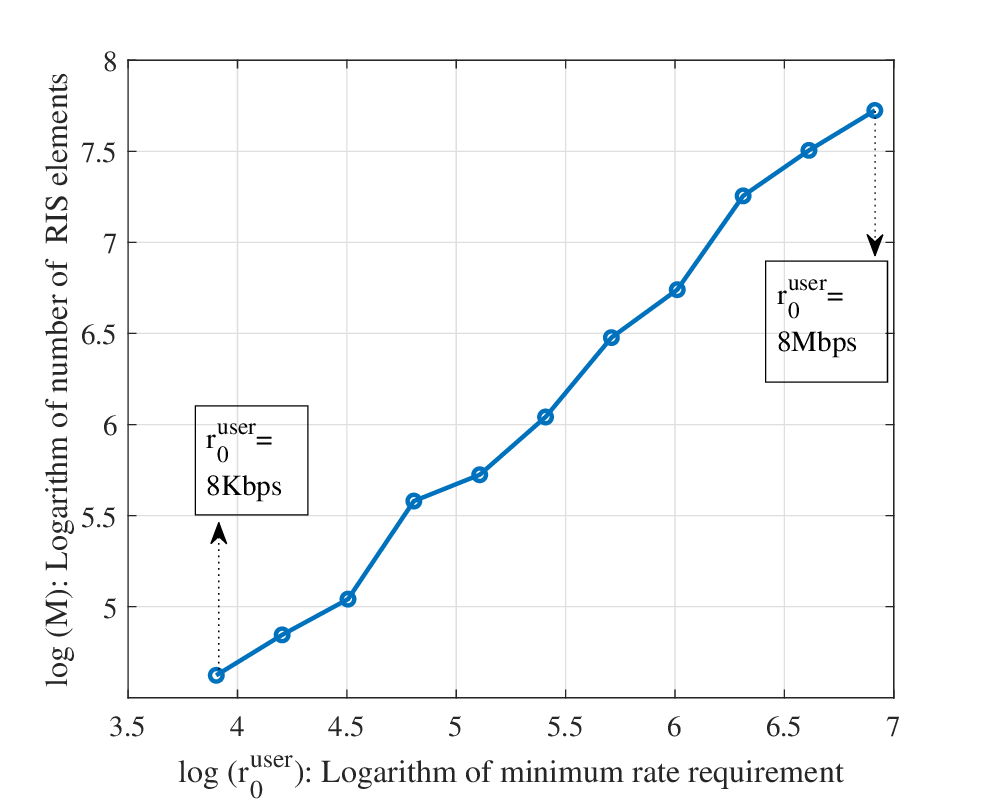}
    \caption{Required number of total RIS elements in logarithmic scale versus logarithm of the minimum rate requirements to provide full coverage, where only HAPS-RIS setup is in use.}
    \label{logMr}
\end{figure}

\textcolor{black}{ Fig. \ref{Me5CoveredUsers}, Fig. \ref{CoveredUsersVSPowerMe5}, and Fig. \ref{logMr} are results obtained by solving the leader problem  \textbf{OP 1} by exploiting the proposed DyRaZARC technique when there is only HAPS-RIS to support the users.} Fig. \ref{Me5CoveredUsers} and Fig. \ref{CoveredUsersVSPowerMe5} are the results where a portion of the coverage area is covered by HAPS-RIS and the rest of the users are in the outage. Fig. \ref{logMr} demonstrates the result where full coverage is provided only by HAPS-RIS system. \textcolor{black}{Fig. \ref{NversusM} is the result obtained by exploiting the proposed DyRaZARC and KADUS techniques to solve the leader and follower problems, i.e., \textbf{OP 1} and \textbf{OP 2}, respectively. In this case, full coverage is provided by integration of two types of infrastructures, i.e., UAVs and HAPS-RIS system.} \textcolor{black}{We take into account a wide range of rate requirements, spanning from kbps to Mbps.}
\subsection{Only HAPS-RIS is in Use}
  Fig. \ref{Me5CoveredUsers} demonstrates the percentage of users covered by HAPS-RIS versus the number of RIS elements for three different data rate thresholds of 64 kbps, 128 kbps, and 256 kbps. As the number of RIS elements increases, the percentage of covered users also rises across all data rate thresholds. For the lowest data rate of 64 kbps, user coverage begins at approximately 50\% and approaches 90\% as the number of RIS elements increases. For the 128 kbps threshold, coverage starts at around 20\% and increases to about 60\%. The highest data rate of 256 kbps shows the lowest coverage, beginning slightly above 10\% and reaching around 30\%. Fig. \ref{CoveredUsersVSPowerMe5} demonstrate the percentage of users covered by HAPS-RIS versus the CS transmit power for various minimum data rate requirements, data rate thresholds of 64 kbps, 128 kbps, and 256 kbps, utilizing $3.5\times10^5$ RIS elements. The findings show that increasing CS transmit power leads to a higher percentage of covered users. These results underscore the pivotal role of both CS transmit power and the number of RIS elements in the coverage provided by HAPS-RIS.
  Fig. \ref{logMr} shows that as the minimum rate requirement \(r^{\rm{user} }_{0} \) increases, the number of RIS elements \( M \) needed for full coverage also rises, both on a logarithmic scale. This indicates that higher rate demands require more RIS elements. The result highlights the potential of the HAPS-RIS system to provide full coverage while satisfying different rate requirements without relying on UAVs.
  
\subsection{HAPS-RIS and UAVs are in Use}
\begin{figure}
\centering
  \includegraphics[scale=0.4]{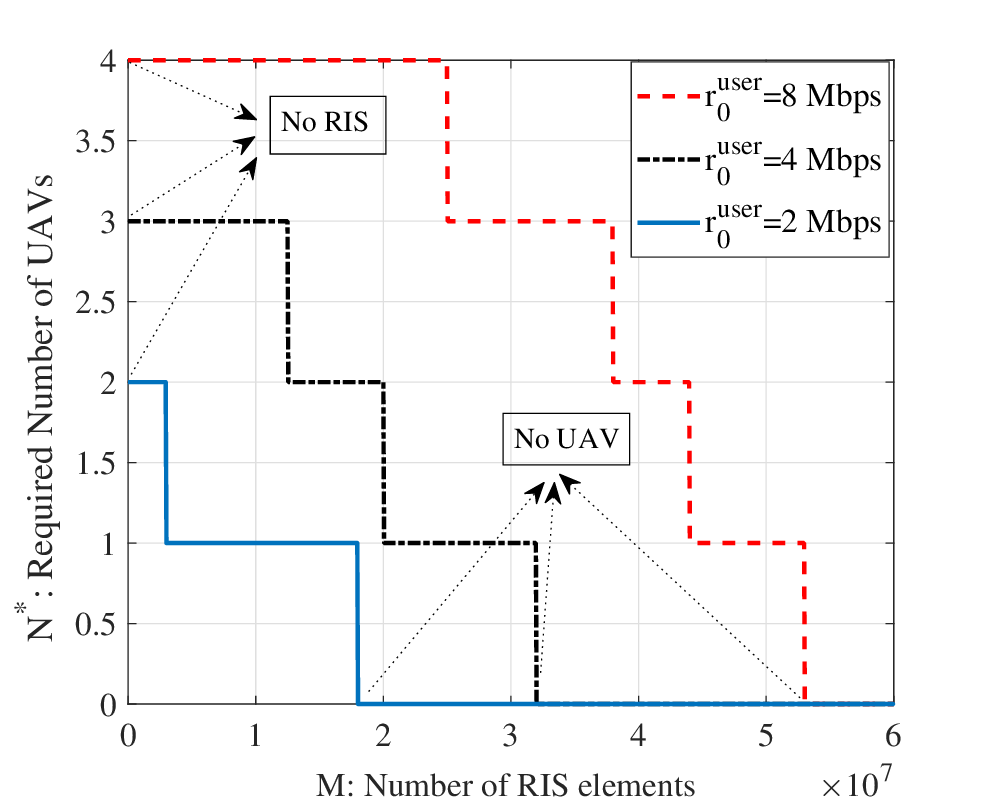}
    \caption{Trade-off between the number of UAVs and RIS elements to provide the full coverage. \textcolor{black}{This is achieved by exploiting the proposed DyRaZARC and KADUS techniques where both HAPS-RIS setup and UAVs are in use.}}
    \label{NversusM}
\end{figure}

Fig. \ref{NversusM} presents the minimum number of required UAVs \( N^{*} \)  versus the total number of RIS elements \( M \) to achieve the full coverage at different minimum rate requirements \(r^{\rm{user} }_{0} \). The graph depicts three distinct scenarios: \(r^{\rm{user} }_{0} = 2 \) Mbps, \(r^{\rm{user} }_{0} = 4 \) Mbps, and \(r^{\rm{user} }_{0} = 8 \) Mbps. The stepped nature of each line indicates that as the number of RIS elements increases, the number of UAVs needed decreases in discrete steps. For the highest rate requirement, 8 Mbps, the system requires up to 4 UAVs when the number of RIS elements is low; however, as \( M \) increases beyond approximately 5 million elements, no UAVs are needed. Conversely, for lower rate requirements, 2 Mbps and 4 Mbps, fewer UAVs are required overall, and the need for UAVs is eliminated at a lower threshold of RIS elements. This analysis illustrates the trade-off between the number of RIS elements and UAVs: as the number of RIS elements increases, the dependence on UAVs decreases, which is particularly beneficial for meeting higher rate requirements with fewer UAVs.



\section{Conclusion}
\textcolor{black}{This paper proposes a HAPS-RIS-assisted network architecture to enhance UAV-based wireless coverage, addressing the challenge of limited UAV availability. A hierarchical bilevel optimization framework is designed, where the leader problem determines the zone boundary and practical RIS phase shifts to maximize HAPS-RIS-supported users under rate constraints. This is achieved using a relaxed phase design and the proposed DyRaZARC technique. The follower problem minimizes the number of UAVs required to cover remaining users by applying the proposed geometry-informed machine learning KADUS technique, which adapts $k$-means clustering to communication constraints and geometric initialization.
Our study reveals that increasing the number of RIS elements significantly reduces the number of required UAVs, enables full user coverage even at high data rates, and supports scalable NTN deployment. This study opens several avenues for future research. One direction is to extend the model to mobile UAV trajectories with energy and path planning constraints. Another is to integrate sensing capabilities into the proposed network architecture, enabling intelligent, context-aware NTN deployment through integrated sensing and communication (ISAC).}



\end{document}